\documentclass[]{spie}  
\usepackage[]{aas_macros}
\usepackage{wrapfig}

\usepackage{amsmath,amsfonts,amssymb}
\usepackage{graphicx}
\usepackage[colorlinks=true, allcolors=blue]{hyperref}

\def\ra{\mathrm{a}}

\newcommand{\rd}{\mathrm{d}}

\def\bH{{\bf H}}

\def\rk{\mathrm{k}}

\def\rm{\mathrm{m}}

\def\rp{\mathrm{p}}

\def\br{\boldsymbol{r}}
\def\rr{\mathrm{r}}

\def\ru{\mathrm{u}}

\def\bx{{\bf x}}
\def\by{{\bf y}}

\def\balpha{\boldsymbol{\alpha}}

\def\brho{\boldsymbol{\rho}}

\title{Empirical Green's Function Approach for Utilizing Millisecond Focal and Pupil Plane Telemetry in Exoplanet Imaging}

\author[*]{Richard A. Frazin}
\affil[*]{Dept. of Climate and Space Sciences, University of Michigan, Ann Arbor, MI 48109}

\authorinfo{Email: rfrazin {\it at} umich.edu}

\begin{document} 
\maketitle

\begin{abstract}

Millisecond focal plane telemetry is now becoming practical due to a new generation of near-IR detector arrays with sub-electron noise that are capable of kHz readout rates.
Combining these data with those simultaneously available from the wavefront sensing system allows the possibility of self-consistently determining the optical aberrations (the cause of quasi-static speckles) and the planetary image.   
This approach may be especially advantageous for finding planets within about 3 $\lambda / D$\ of the star where differential imaging is ineffective.
As shown in a recent article by the author (J. Opt. Soc. Am. A., 33, 712, 2016), one must account for unknown aberrations in several non-conjugate planes of the optical system, which, in turn,  requires ability to computational propagate the field between these planes.
These computations are likely to be difficult to implement and expensive.   
Here, a far more convenient alternative based on empirical Green's functions is provided.  
It is shown that the empirical Green's function (EGF), which accounts for all multi-planar, non-common path aberrations, and results in a much more tractable and highly parallel computational problem. 
It is also shown that the EGF can be generalized to treat polarization, resulting in the empirical Green's tensor (EGT).

\end{abstract}


\section{INTRODUCTION}\label{Intro}

Direct imaging of exoplanetary systems is difficult due to the high contrast in brightness between the planet and the star, resulting in a situation in which the planetary image is confounded with the transient details of the telescope's point spread function (PSF).
To date, all methods for direct imaging of exoplanets have relied on subtracting the PSF from the observed image to obtain the final science image.
Typically, the PSF subtraction is achieved through differential imaging methods, which have been applied extensively in the recent literature.
The fundamental difficulties associated with differential imaging methods, most notably angular differential imaging (ADI) and spectral differential imaging (SDI), have been reviewed in [\citenum{Frazin13,Frazin14,Frazin16a,Frazin16b,Marois_SOSIE,Rameau_ADI_SDI_limits15}] and references therein.  
Briefly, SDI is nearly useless close to the host star because the difference in the point spread function (PSF) at two wavelengths is proportional to the distance from the center of the image, and ADI assumes that the optical aberrations are not changing in time.
Since differential imaging methods require estimating the speckle background from the final image, they suffer from a severe statistical penalty close to center \cite{Mawet_SmallNumStatsSpeckle14}, which is the most scientifically fruitful part of the image \cite{Stark_ExoEarthYield14,Brown_PlanetSearch15}.

It is the author's belief that the combination of millisecond pupil plane and focal plane telemetry will lead to a large improvement in contrast over differential imaging techniques, as it leverages a vastly larger and richer data set than methods that use standard exposure times, which average over the atmospheric turbulence.
Such methods are becoming practical due to a new generation of ultra-low noise IR cameras capable of kHz readouts, such as the SWIR single photon detector, SAPHIRA  eAPD and the MKIDS \cite{SWIR_detector14,SELEX_APD12,Saphira_eAPD14,Mazin_MKIDS14}.
It is important to emphasize that the author's millisecond imaging techniques can be generalized to take advantage of essentially all constraints on problem proposed to date.   
These constraints include those imposed by diurnal rotation (used by ADI), multi-wavelength observations (used by SDI), as well polarization (used in polarization differential imaging \cite{Hinkley_PDI09}).
The basic concept of the author's approach was explained in [\citenum{Frazin13,Frazin14}].
Essentially, these papers provide a system of equations that express the instantaneous image measured in the science camera (SC) in terms of:
\begin{itemize}
\item{the residual phase [which is measured by the wavefront sensor (WFS)]}
\item{the planetary image}
\item{non-common path aberrations (NCPA).}
\end{itemize}

\begin{wrapfigure}{r}{0pt}
\includegraphics[width=.49\linewidth,clip=]{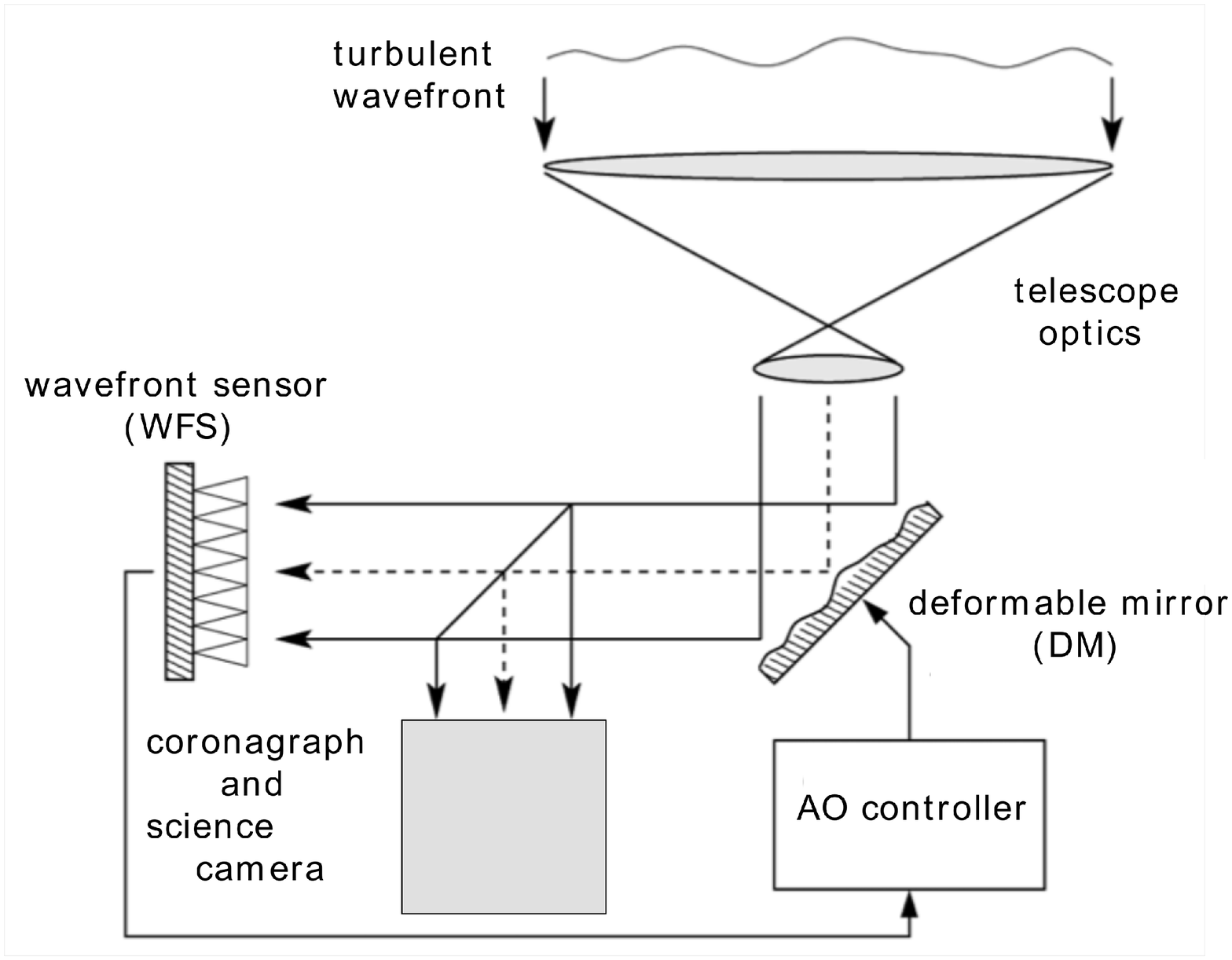}
\caption{\small Schematic diagram of an astronomical telescope with a closed-loop AO system and a coronagraph.  Modified from [\citenum{Hinnen_H2control}].}
\label{fig_schematic}
\vspace{-4mm}
\end{wrapfigure}

\noindent This system of equations forms the basis of a regression problem in which the NCPA and the planetary image are estimated self-consistently.   
The information required to successfully perform the estimation is provided by the variability of the residual phase (so it is closely related to other phase diversity methods), although multi-wavelength and diurnal rotation constraints could be employed as well. 
In [\citenum{Frazin16a,Frazin16b}], the author extended this approach to include non-pupil plane NCPA, which requires summing the contributions from unknown aberrations in various optical planes.
The resulting expression is very complicated and computationally demanding because it requires propagation operators between the various optical planes.   
Here, the problem is formulated in full rigor, but in a way that is far more computationally tractable and should be suitable for parallel computation.
The disadvantage of this approach is that the model has more degrees of freedom, but the richness of the millisecond data streams may well afford such a luxury.

\section{Image Formation Equations}
\subsection{Optical Propagation}

Consider a telescope with a single conjugate, closed-loop AO system and a coronagraph, similar to that shown in Fig.~\ref{fig_schematic}.
Assume that the AO system is using the only the star at the center of the putative solar system under investigation to track the turbulence.
As per the discussion in [\citenum{Frazin16a}], the star at the center of the putative planetary system gives rise to an electromagnetic disturbance impinging on the telescope entrance pupil that can be represented as:
\begin{equation}
u_0(\br_0,t) = \sqrt{I_\star \,} \exp j \big[  k \balpha_\star \cdot \br_0 + \phi_\ra(\br_0,t) \big]  \, ,
\label{u_0}
\end{equation}
where $u_0(\br_0,t)$\ is the complex-valued electric field impinging on the telescope entrance aperture (represented by index value $0$), $t$\ represents time, $\br_0$, which has units of distance, is the two-dimensional (2D) coordinate in the telescope entrance pupil plane, $\balpha_\star$, which has units of radians, is the (small) 2D sky-angle of the star relative to the telescope pointing direction (pointing error), and $k= 2 \pi / \lambda $\ is the wavenumber corresponding to wavelength $\lambda$, $\phi_\ra(\br_0,t)$\ is the complex-valued atmospheric modulation, and $I_\star$\ is the intensity of the starlight. 

Again considering Fig.~\ref{fig_schematic}, the stellar field incident on the deformable mirror (DM) surface, denoted with index $D$, is given by:
\begin{equation}
u_D(\br_D,t) 
 = \sqrt{I_\star} \Upsilon_{D,0}(\br_D,\br_0)
\exp j \big[ k \balpha_\star \cdot \br_0 + \phi_\ra(\br_0,t)  \big]  \, ,
\label{u_D-} 
\end{equation}
where $\Upsilon_{D,0}$\ is a propagation operator that relates the field at the telescope entrance (plane $0$) to the field at the DM (plane $D$), and $\br_D$\ is the coordinate on the DM surface.
Under the operator notation established in [\citenum{Frazin16a}], the $\Upsilon_{D,0}(\br_D,\br_0)$\ operator includes 2D integration over the entrance pupil coordinate $\br_0$.
In keeping with notational conventions established in [\citenum{Frazin16a}], the operator $\Upsilon_{B,A}$, which propagates the field from plane $B$ to plane $A$, includes interaction with the optical surface $A$, but does not include interaction with optical surface $B$.

As can be seen in Fig.~\ref{fig_schematic}, after interacting with the DM the light goes through a beam splitter (BS), with some light going to the WFS and some to the coronagraph and SC.  
For optical systems with a different design, e.g., using dichroic instead of a BS, let BS represent the plane in which an optical element divides the light between WFS and SC arms of the optical system.
Thus, the BS represents the last place where the where the light in the WFS and SC experience a common optical path.
Let the BS, WFS, SC detector planes be denoted with the indices $B$, $w$ and $C$, respectively.
Then, the WFS and SC fields arising from the star can be expressed as:
\begin{align}
u_w(\br_w,t) &= \sqrt{I_\star \,} \Upsilon_{w,B}(\br_w,\br_B) \Upsilon_{B,D}(\br_B,\br_D) 
 \Upsilon_{D,0}(\br_D,\br_0)  
\exp  j \big[ k \alpha_\star \cdot \br_0 +   \phi_\ra(\br_0,t) \big]  
\label{u_w_complete}\\ 
 u_C(\br_C,t) & = \sqrt{I_\star \,} \Upsilon_{C,B}(\br_C,\br_B) \Upsilon_{B,D}(\br_B,\br_D) 
  \Upsilon_{D,0}(\br_D,\br_0)  
\exp  j \big[ k \alpha_\star \cdot \br_0 +   \phi_\ra(\br_0,t)  \big] \, .
\label{u_C_complete}
\end{align}
Note that the operator $\Upsilon_{B,D}$\ includes the effect of the DM.
Eq.~(\ref{u_w_complete}) assumes that the planetary emission makes only a negligible contribution to the WFS signal (which is undoubtedly the case in high-constrast applications), but this is not an issue for Eq.~(\ref{u_C_complete}) since the planetary contribution to the SC signal will be added later.
Note that the only difference in these two expressions is the leftmost operator, which propagates the light from the BS to the WFS in Eq.~(\ref{u_w_complete}), and from the BS to the SC in Eq.~(\ref{u_C_complete}).
If the WFS is to provide useful information about the light in telescope entrance pupil, which it must do for the AO system to work, the operator $ \Upsilon_{w,B}$\ must be invertible or nearly so (meaning it does not destroy much information), allowing an approximate inverse.  
Let the appropriate inverse operator be denoted as $\Upsilon^{-1}_{w,B}(\br_B,\br_w)$ (noting the order of the arguments), which calculates the field in the BS plane ($B$) from the field in the WFS plane ($w$).
Placing $\Upsilon^{-1}_{w,B}(\br_B,\br_w)$\ on the left of both sides of Eq.~(\ref{u_w_complete}), it is easy to see that Eq.~(\ref{u_C_complete}) can be rewritten as:
\begin{align}
 u_C(\br_C,t) & = \Upsilon_{C,B}(\br_C,\br_B) \Upsilon^{-1}_{w,B}(\br_B,\br_w) u_w(\br_w,t)\,  \label{u_C-u_w-1} \\
 & = \Upsilon_{C,w}(\br_C,\br_w)  u_w(\br_w,t)\,  \label{u_C-u_w} \, , 
\end{align}
where new operator $\Upsilon_{C,w}$\ is the contraction of $\Upsilon_{C,B}$\ and $ \Upsilon^{-1}_{w,B}$, i.e.,
\begin{equation}
\Upsilon_{C,w} \big( \br_C,\br_w) \equiv  \Upsilon_{C,B}(\br_C,\br_B) \Upsilon^{-1}_{w,B}(\br_B,\br_w) \, ,
\label{Upsilon_{Cw}-def}
\end{equation}
in which the integration over the intermediate coordinate $\br_B$ is implicit.
Eq.~(\ref{u_C-u_w}) is fundamental to this analysis because it relates the SC field to the WFS field.

On the other hand, if we define the residual phase $\phi_\rr(\br_w,t)$\ to be (complex-valued) phase of the starlight impinging on the WFS (including the effects of NCPA), then the field impinging on the WFS is also given by:
\begin{equation}
u_w(\br_w,t) = \sqrt{I_\star} \exp j [ \beta k \balpha_\star \cdot \br_w +  \phi_\rr(\br_w,t) ] \, , 
\label{u_w-resid}
\end{equation}
where the pointing error $\balpha_\star$\ is included explicitly, and the factor $\beta$\ is the demagnification.
For example, in an 8 m telescope with a 2 cm WFS entrance pupil, the $\beta = [8 \, \mathrm{m}] / [2 \, \mathrm{cm} ] = 400$. 

\subsection{Empirical Green's Function (EGF)}

In [\citenum{Frazin16b}], the operators $\Upsilon_{C,B}$ and $ \Upsilon^{-1}_{w,B}$ from Eq.~(\ref{u_C-u_w-1}) are then expressed in terms of unknown aberration functions in various planes and propagation operators between those planes, resulting in a sum that potentially requires a large number of propagation operations.   
In order to avoid such difficult computation, an alternative approach in terms of an empirical Green's function (EGF) is introduced here.   
The EGF method is easier to implement and relies on somewhat weaker assumptions than expressing the stellar field explicitly in terms of the unknown aberrations in multiple planes.  
The disadvantage of the EGF method is that it likely implies more unknown parameters than the aberration expansion, however, the resulting EGF computation has a much more computationally convenient block-diagonal structure that is suitable to parallel computation.

The starting point for the EGF is Eq.~(\ref{u_C-u_w-1}), and the first step is to express $ \Upsilon^{-1}_{w,B}(\br_B,\br_w) $\ as:
\begin{equation}
 \Upsilon^{-1}_{w,B}(\br_B,\br_w) =  \Upsilon^{\rk-1}_{w,B}(\br_B,\br_w) + \Upsilon^{\ru-1}_{w,B}(\br_B,\br_w) \, ,
\label{EGF-wC}
\end{equation}
where $ \Upsilon^{\rk-1}_{w,B}(\br_B,\br_w)$\ contains what is known about the true operator $ \Upsilon^{\rk-1}_{w,B}(\br_B,\br_w)$\ and $ \Upsilon^{\ru-1}_{w,B}(\br_B,\br_w)$\ is an unspecified propagation operator containing what is not known.  
Therefore it accounts for any sort of NCPA, whether or not is it in the pupil plane.
Similarly, the operator $\Upsilon_{C,B}(\br_C,\br_B)$ can be split into known and unknown parts as 
\begin{equation}
\Upsilon_{C,B} \big( \br_C,\br_B \big) = \Upsilon^\rk_{C,B} \big( \br_C,\br_B \big)  + 
\Upsilon^\ru_{C,B} \big( \br_C,\br_B \big) \, ,
\label{EGF-BC}
\end{equation}
where the operator $\Upsilon^\ru_{C,B} \big( \br_C,\br_B \big)$\ contains what is unknown about the operator $\Upsilon_{C,B} \big( \br_C,\br_B \big)$.
There is, in fact, another advantage to Eq.~(\ref{EGF-BC}) over the aberration expansion: it assumes only linear in the electric field, and does not require the unknown aberrations to be small.
Combining Eqs.~(\ref{EGF-wC}) and (\ref{EGF-BC}) yields
\begin{equation}
\Upsilon_{C,w} \big( \br_c,\br_w \big) = \Upsilon^\rk_{C,w} \big( \br_C,\br_w \big)  + 
\Upsilon^\ru_{C,w} \big( \br_C,\br_w \big) \, ,
\label{EGF-Cw}
\end{equation}
where 
\begin{equation}
\Upsilon^\rk_{C,w} \big( \br_C,\br_w \big) \equiv \Upsilon^\rk_{C,B} \big( \br_C,\br_B \big) \Upsilon^{\rk-1}_{w,B}(\br_B,\br_w)
\label{EGF-Cw-known}
\end{equation}
 is a known operator, e.g., a stellar coronagraph model, and 
\begin{multline}
\Upsilon^\ru_{C,w} \big( \br_C,\br_w \big) \equiv    \Upsilon^\rk_{C,w} \big( \br_C,\br_w \big) \Upsilon^{\ru-1}_{w,B}(\br_B,\br_w) + 
\Upsilon^\ru_{C,B} \big( \br_C,\br_B \big) \Upsilon^{\rk-1}_{w,B}(\br_B,\br_w)  + \\
\Upsilon^\ru_{C,B} \big( \br_C,\br_B \big) \Upsilon^{\ru-1}_{w,B}(\br_B,\br_w) 
\label{EGF-Cw-unknown}
\end{multline}
is an unknown operator.
One key advantage of the EGF formulation is that the operators $ \Upsilon^{\ru-1}_{w,B}$\ and $\Upsilon^\ru_{C,B} $\ never need to be determined, instead it remains to estimate the $\Upsilon^\ru_{C,w} \big( \br_C,\br_w \big) $, which accounts for all NCPA (pupil plane or otherwise) and aberrations downstream of the BS.
The propagation operator $ \Upsilon^\ru_{C,w} \big( \br_C,\br_w \big) $ will be known as the ``empirical Green's function,'' since the Green's function (general solution to the diffraction problem) is the known (modeled) part plus the EGF.
Using Eqs.~(\ref{EGF-Cw}) and (\ref{u_w-resid}), Eq.~(\ref{u_C-u_w})  becomes:
\begin{equation}
 u_C(\br_C,t) = \sqrt{I_\star \,} \big[  \Upsilon^\rk_{C,w} \big( \br_C,\br_w \big)   + \Upsilon^\ru_{C,w} \big( \br_C,\br_w \big)  \big]
\exp j [ \beta k \balpha_\star \cdot \br_w +  \phi_\rr(\br_w,t) ] \, ,
\label{EGF-u_C-u_w}
\end{equation}
Which expresses the SC field in terms of EGF and the field impinging on the WFS.
Eq.~(\ref{EGF-u_C-u_w}) includes the effects of NCPA and aberration downstream of the BS.

The essence of the EGF regression method is to estimate the operator $\Upsilon^\ru_{C,w} \big( \br_C,\br_w \big)$\ (and the planetary image given the SC and WFS data streams.
This requires expressing the intensity measured by the SC in terms of the EGF.
The intensity of the starlight impinging on the SC detector is $I_C(\br_C,t) = u_C(\br_C,t) u_C^*(\br_C,t)$, where the science camera field, $u_C$, is given by Eq.~(\ref{EGF-u_C-u_w}), resulting in:
\begin{multline}
I_{\star C}(\br_C,t) = I_\star \bigg[ 
    \Upsilon^\rk_{C,w} \big( \br_C,\br_w \big)   \Upsilon^{\rk *}_{C,w} \big( \br_C,\br_w' \big)  
+  \Upsilon^\rk_{C,w} \big( \br_C,\br_w \big) \Upsilon^{\ru *}_{C,w} \big( \br_C,\br_w' \big)
+  \Upsilon^{\rk *}_{C,w} \big( \br_C,\br_w \big)  \Upsilon^\ru_{C,w} \big( \br_C,\br_w' \big) \\
+  \Upsilon^{\ru}_{C,w} \big( \br_C,\br_w \big) \Upsilon^{\ru *}_{C,w} \big( \br_C,\br_w' \big)
\bigg] 
 \exp j \big[ \beta k \balpha_\star \cdot (\br_w - \br_w') + 
 \phi_\rr(\br_w,t) -  \phi_\rr^*(\br_w',t)  \big] \, .
\label{EGF-SC-coherency-scalar}
\end{multline}
The first term in brackets in Eq.~(\ref{EGF-SC-coherency-scalar}) contains only the known model of the optical system.
The second and third term are linear in the (unknown) EGF $ \Upsilon^\ru_{C,w} $\, while the fourth term is quadratic in the EGF.   
Below, it is shown that (\ref{EGF-SC-coherency-scalar}) result in a block-diagonal system of equations that should be well suited to parallel computation.

\subsubsection*{Phase Sorting Interferometry}

This section relates EGF presented here to ``phase sorting interferometry'' method of [\citenum{Codona13}].
Assume that $| \phi_\rr(\br_w,t)|$ is small enough so that $\exp \big [ j  \phi_\rr(\br_w,t) \big] \approx 1 +  j\phi_\rr(\br_w,t) $\ and the pointing error $\balpha_\star = 0$.
Then, Eq.~(\ref{EGF-u_C-u_w}) becomes:
\begin{equation}
 u_C(\br_C,t) \approx \sqrt{I_\star \,} \left\{ \Upsilon^\rk_{C,w} \big( \br_C,\br_w \big) \exp j [  \phi_\rr(\br_w,t) ]  
+ \Upsilon^\ru_{C,w} \big( \br_C,\br_w \big) \big[ 1 +  j \phi_\rr(\br_w,t) \big] \right\} \, .
\label{Codona}
\end{equation}
Note that the second term, which is just the EGF itself, $\Upsilon^\ru_{C,w} \big( \br_C,\br_w \big)$\ integrates over $\br_w$, and is only a function of the image plane coordinate $\br_C$.
This is the is the unknown ``subject field'' ($\Psi$\ in the notation of of [\citenum{Codona13}]), which corresponds to the quasi-static speckle that one wants to subtract and their method seeks to determine.
The analysis of [\citenum{Codona13}] does not include the third term $j \Upsilon^\ru_{C,w} \big( \br_C,\br_w \big) \phi_\rr(\br_w,t) $.

\section{Block-Diagonal Form of the EGF Regression}

The EGF regression is block-diagonal because there is a separate and statistically independent regression problem associated with each spatial pixel in the SC.  
Thus, each ``block'' corresponds to a SC pixel (and the time-evolution of its intensity).
This is demonstrated below.

First, define $I_0(\brho_l,t_i)$, to be the intensity of the starlight impinging on the SC in the absence of unknown aberration, i.e.,  that which would arise if the EGF were 0.
It is given by the expression:
\begin{equation}
I_0(\brho_l,t_i) \equiv \\ I_\star \Upsilon^\rk _{C,w} (\brho_l,\br) \Upsilon^{\rk*}_{C,w}(\brho_l,\br') 
 \exp j \big[ \beta k \balpha_\star \cdot (\br - \br') + \phi_\rr(\br,t_i) -  \phi_\rr^*(\br',t_i) \big] \, ,
\label{I_null-EGF}
\end{equation}
where $\brho_l \equiv \br_C(\mathrm{pixel \; index}\; l)$\ to be the position of pixel $l$ and $t_i$\ is a timestamp from the WFS data stream.
Note that the final term of Eq.~(\ref{EGF-SC-coherency-scalar}),$ \Upsilon^{\ru}_{C,w} \big( \brho_l,\br \big) \Upsilon^{\ru *}_{C,w} \big( \brho_l ,\br \big)$, is quadratic in the EGF and presumably smaller than the  sum $\Upsilon^\rk_{C,w} \big( \brho_l,\br \big) \Upsilon^{\ru *}_{C,w} \big( \brho_l,\br \big) +  \Upsilon^{\rk *}_{C,w} \big( \brho_l,\br \big)  \Upsilon^\ru_{C,w} \big( \brho_l,\br \big)$.   
Thus, there is justification for dropping it.  Even if it not necessarily negligible it can be linearized and treated iteratively, which is straightforward.
Dropping the quadratic term, Eq.~(\ref{EGF-SC-coherency-scalar}) becomes:
\begin{multline}
I_{\star C}(\brho_l,t_i) - I_0(\brho_l, t_i)  \approx \\
\big[ \Upsilon^\rk_{C,w} \big( \brho_l,\br \big) \Upsilon^{\ru *}_{C,w} \big( \brho_l,\br' \big)
+  \Upsilon^{\rk *}_{C,w} \big( \brho_l,\br \big)  \Upsilon^\ru_{C,w} \big( \brho_l,\br' \big) \big]
 \exp j \big[ \beta k \balpha_\star \cdot (\br - \br') + 
 \phi_\rr(\br,t) -  \phi_\rr^*(\br',t) \big] \, ,
\label{EGF-approx}
\end{multline}
where the factor $\sqrt{I_\star}$\ has been absorbed into the operators.
For a fixed value of $\brho_l$\ the EGF $\Upsilon^\ru_{C,w} \big( \brho_l,\br \big)$\ can be considered to be a function of the coordinate $\br$\ and can be represented with series expansion:
\begin{equation}
 \Upsilon^\ru_{C,w} \big( \brho_l,\br \big) = \sum_{k=0}^{N_l-1} a_{l,k} \psi_{l,k}(\br) \, , 
\label{EGF-expansion}
\end{equation}   
where the $\{ \psi_{l,k}(\br) \} $\ are a set of pre-specified expansion functions (e.g., annular Zernike polynomials) and the $\{ a_{l,k} \}$\ are unknown expansion coefficients. 
Eq.~(\ref{EGF-expansion}) has the very desirable effect of creating a block-diagonal regression problem, in which one obtains an independent set of equations at each SC pixel position $\brho_l$, as will be made clear below.
The number of terms in the expansion $N_l$\ and the choice of expansion functions $\{ \psi_{l,k}(\br) \} $\ can different for each SC position $\brho_l$, allowing considerable flexibility.
Note that there is no requirement that the $\{ \psi_{l,k}(\br) \} $\ satisfy orthogonality conditions.
The terms $\{ a_{l,k} \psi_{l,k}(\br) \}$\ must be complex-valued, so that the $\{ a_{l,k} \}$\ must be complex-valued if the $\{ \psi_{l,k}(\br) \} $\ are taken to be real, as they will be here.
To make this explicit, define $a_{l,k} \equiv a_{l,k}^r + j a_{l,k}^i$, splitting it into real and imaginary parts.  
Then Eq.~(\ref{EGF-approx}) becomes
\begin{equation}
I_{\star C}(\brho_l,t_i) - I_0(\brho_l, t_i)  \approx   \sum_{k=0}^{N_l-1} a_{l,k}^r h_{l,k,i}^{R}  \; + \;
j \sum_{k=0}^{N_l-1} a_{l,k}^i h_{l,k,i}^{I} \, , 
 \label{EGF-approx-expan}
\end{equation}
where
\begin{align}
   h_{l,k,i}^{R} & \equiv \big[  \Upsilon^\rk_{C,w} \big( \brho_l,\br \big)  \psi_{l,k}(\br') + \Upsilon^{\rk *}_{C,w} \big( \brho_l,\br' \big)  \psi_{l,k}(\br)   \big]  \exp j \big[ \beta k \balpha_\star \cdot (\br - \br') + 
 \phi_\rr(\br,t) -  \phi_\rr^*(\br',t) \big]  \: \: \: \mathrm{and} \nonumber \\
   h_{l,k,i}^{I}  & \equiv \big[  \Upsilon^{\rk *}_{C,w} \big( \brho_l,\br' \big)  \psi_{l,k}(\br) - \Upsilon^{\rk }_{C,w} \big( \brho_l,\br \big)  \psi_{l,k}(\br')   \big] \exp j \big[ \beta k \balpha_\star \cdot (\br - \br') + 
 \phi_\rr(\br,t) -  \phi_\rr^*(\br',t) \big]  \, .
 \label{hR-hI-defs}
\end{align}

The measured image in the SC, denoted by $I_\rm(\brho_l,t_i)$, has an additive random component $\nu(\brho_l,t_i)$, which results from shot noise and detector readout noise (if nothing else), and is given by: 
\begin{equation}
I_\rm(\brho_l,t_i) = I_{\star C}(\brho_l,t_i) + \nu(\brho_l,t_i) \, .
\label{I_measured}
\end{equation} 
Assuming that $\nu(\brho_l,t_i)$\ and $\nu(\brho_{l'},t_i)$\ are statistically independent (which, hopefully, is a good approximation for most readout noise processes), one obtains the following regression problem in a canonical form, separately for each pixel index $l$:
\begin{equation}
\by_l = \bH_l \bx_l \, ,
\label{Y=Hx}
\end{equation}
where 
 \begin{equation}
 \by_l = 
\left[
\begin{array}{c} 
I_\rm(\brho_l,t_0) - I_0(\brho_l,t_0) - \nu(\brho_l,t_0) \\
\vdots \\
I_\rm(\brho_l,t_i) - I_0(\brho_l,t_i) -  \nu(\brho_l,t_i)  \\
\vdots \\
I_\rm(\brho_l,t_{T-1}) - I_0(\brho_l,t_{T-1}) -  \nu(\brho_l,t_{N_i - 1}) 
\end{array}  
 \right] \, ,
 \label{y_l}
\end{equation}
 \begin{equation}
 \bx_l = 
\left[
\begin{array}{c} 
a^r_{l,0} \\
\vdots \\
a^i_{l,N_l-1}  \\
a^r_{l,0} \\
\vdots \\
a^i_{l,N_l-1}  \\
\end{array}  
 \right] \, , \: \: \mathrm{and}
\label{x}
\end{equation}
\begin{equation}
 \bH_l = 
\left[ \begin{array}{l l l l l l l l l l} 
h_{l,0,0}^R & \hdots & h_{l,k,0}^R & \hdots & h_{l,N_l-1,0}^R & h_{l,0,0}^I & \hdots & h_{l,k,0}^I & \hdots & h_{l,N_l-1,0}^I \\
\vdots    &             & \vdots    &            & \vdots          & \vdots    &             & \vdots    &             & \vdots \\
h_{l,0,i}^R & \hdots & h_{l,k,i}^R & \hdots & h_{l,N_l-1,i}^R & h_{l,0,i}^I & \hdots & h_{l,k,i}^I & \hdots & h_{l,N_l-1,i}^I \\
\vdots    &             & \vdots    &            & \vdots          & \vdots    &             & \vdots    &             & \vdots \\
h_{l,0,T-1}^R & \hdots & h_{l,k,T-1}^R & \hdots & h_{l,N_l-1,T-1}^R & h_{l,0,T-1}^I & \hdots & h_{l,k,T-1}^I & \hdots & h_{l,N_l-1,T-1}^I \\
\end{array} \right] \, ,
 \label{y_l}
\end{equation}
 where $T$\ is the total number of exposures to be included in the regression.
The goal of the regression is to estimate $\bx_l$, which contains the coefficients that describe the EGF.

Thus, for each SC pixel, given by the index $l$, one estimates the $2N_l$\ expansion coefficients contained in the vector $\bx_l$.  
It may be practical to solve for a sufficient number of coefficients to allow a rather high-order expansion in Eq.~(\ref{EGF-approx-expan}).
As an example, assume the expansion functions $\{ \psi_{l,k}(\br) \}$\ are given by the annular Zernike polynomials up to 20\underline{th} order, which would correspond to a total of $230$\ polynomials, or $N_l = 230$.   
Since we must solve for the real and imaginary parts of the expansion coefficients, the vector $\bx_l$ would have $460$ components that must be estimated from the regression.   
Further assume that the cadence of the SC and WFS is 1 millisecond, the atmospheric clearing time (the time it take for the wind to cross the telescope) is 2 seconds, and that we collect data for 10 clearing times (20 seconds).
Then, in each SC pixel there are $2\times 10^4$\ observations from which to estimate these 460 coefficients, overdetermining the problem by a factor of about 40.   
The number of coefficients that one may determine is likely to be limited by the inversion of a matrix of size $2 N_l \times 2 N_l $ (computation time proportional to $N_l^3$), or $460 \times 460$\ in this example.   Note that the author's desktop machine was able to invert a $1000 \times 1000$ matrix of random numbers (which tends to result in a poorly conditioned system) in about 0.1 s.

\subsection*{Including the Planetary Image}

The regression problem in Eq.~(\ref{Y=Hx}) does not include the planetary image.
In order to include the planetary image, its intensity must be added to Eq.~(\ref{I_measured}). 
The intensity arising on the SC due to planetary emission, denoted by $I_{\rp C}$, is given by [\citenum{Frazin13,Frazin16a,Frazin16b}]:
\begin{equation}
I_{\rp C}(\brho_l,t_i) = \int_\rp   \rd \balpha \, S(\balpha) \bigg\{ 
 \Upsilon_\rp(\brho_l,\br)   \Upsilon^*_\rp(\brho_l,\br') 
\exp j \big[ \phi_\rr(\br,t_i) - \phi^*_\rr(\br',t_i)  +
 k \beta \balpha \cdot (  \br - \br'   ) \big]  \bigg\} \, ,
\label{pl-SC-coherency-scalar}
\end{equation}
where $\balpha$\ is the 2D sky-angle in units of radians, $\Upsilon_\rp(\brho_l,\br) $\ is a simplified model of the optical system that is suitable calculating the planetary contribution,
 and $S(\balpha)$\ is the sought-after planetary image.
The operator $\Upsilon_\rp(\brho_l,\br) $\ calculates the SC field given the field in some pupil plane of the instrument, quite possibly the DM plane, and the factor $\beta$\ is the demagnification defined earlier.
Eq.~(\ref{pl-SC-coherency-scalar}) assumes that the effects of anisoplanatism are negligible, which should be valid for exoplanet imaging.\cite{Clenet_aniso15,Frazin16b}

The quantity in braces in Eq.~(\ref{pl-SC-coherency-scalar}) the ``planetary intensity kernel.'' 
As described in the author's previous papers, Eq.~(\ref{pl-SC-coherency-scalar}) corresponds to a multi-frame deconvolution problem, with a known PSF.
Estimating the planetary image $S(\balpha)$\ from Eq.~(\ref{pl-SC-coherency-scalar}) couples the spatial pixels in the SC, and therefore does not admit the block-diagonal form of Eq.~(\ref{Y=Hx}).
However, alternating solution methods, in which iteratively updates the estimates of EGF, i.e., $\bx_l$, while holding $S(\balpha)$\ constant, and vice-versa, can utilize the block-diagonal form of Eq.~(\ref{Y=Hx}).

\section{Treating Polarization Effects}

In [\citenum{Frazin16a,Frazin16b}], the author presents a fully polarimetric discussion, allowing for the treatment of polarizing optical elements, the importance of which was demonstrated in [\citenum{Breckinridge15}].
The EGF method can be generalized to treat unknown polarization effects.
The generalization of the EGF to treat polarization is called in ``empirical Green's tensor'' (EGT).  
While the EGF is a scalar operator, the EGT is $2 \times 2$\ matrix of scalar operators that operates on the 2D field described by the Jones vector. 
The vector counterpart of Eq.~(\ref{EGF-u_C-u_w}) is:
\begin{equation}
 u_C(\br_C,t) = \sqrt{I_\star \,} \big[  \Upsilon^\rk_{C,w} \big( \br_C,\br_w \big)   + \Upsilon^\ru_{C,w} \big( \br_C,\br_w \big)  \big]
\breve{u}_w(\br_w,t) 
\exp j [ \beta k \balpha_\star \cdot \br_w +  \phi_\rr(\br_w,t) ] \, ,
\label{EGT-u_C-u_w}
\end{equation}
where the field $u_C$\ is a $2\times 1$ Jones vector, $ \Upsilon^\rk_{C,w}$ is now a $2 \times 2$\ matrix, and  $\breve{u}_w $\ is the $2 \times 1$ polarization state vector of the light impinging on the WFS.\cite{Frazin16b}

When treating polarization effects, one must consider the $4 \times 1$  coherency vector (which differs from the better-known Stokes vector by a linear transformation), instead of the simple scalar intensity.
The coherency vector of the starlight impinging on the SC detector is $J_C(\br_C,t) = u_C(\br_C,t) \otimes u_C^*(\br_C,t)$,\cite{Frazin16a} where $\otimes$\ denotes the Kronecker product, and the science camera field, $u_C$, is given by Eq.~(\ref{EGT-u_C-u_w}).
Expressing the star's coherency in terms of the EGT with  Eq.~(\ref{EGT-u_C-u_w}) results in:
\begin{multline}
J_{\star C}(\br_C,t) = I_\star \bigg[ 
    \Upsilon^\rk_{C,w} \big( \br_C,\br_w \big)  \otimes \Upsilon^{\rk *}_{C,w} \big( \br_C,\br_w' \big)  
+  \Upsilon^\rk_{C,w} \big( \br_C,\br_w \big) \otimes \Upsilon^{\ru *}_{C,w} \big( \br_C,\br_w' \big) \\
+  \Upsilon^{\rk *}_{C,w} \big( \br_C,\br_w \big) \otimes \Upsilon^\ru_{C,w} \big( \br_C,\br_w' \big) 
+  \Upsilon^{\ru}_{C,w} \big( \br_C,\br_w \big) \otimes \Upsilon^{\ru *}_{C,w} \big( \br_C,\br_w' \big)
\bigg] \big( \breve{u}_w(\br_w,t) \otimes \breve{u}_w^*(\br_w',t) \big) \\ \times
 \exp j \big[ \beta k \balpha_\star \cdot (\br_w - \br_w') + 
 \phi_\rr(\br_w,t) -  \phi_\rr^*(\br_w',t)  \big] \, ,
\label{EGT-SC-coherency}
\end{multline}
which makes no approximations that are not implicit in Eq.~(\ref{EGT-u_C-u_w}).
Eq.~(\ref{EGT-SC-coherency}) is the vector  counterpart of  Eq.~(\ref{EGF-SC-coherency-scalar}).

As before, the unknown aspects of the optical system are treating by estimating the EGT, $ \Upsilon^{\ru}_{C,w} $.
Since the EGT is a matrix containing 4 scalar operators, estimating it is more demanding than estimating the EGF, which is a single scalar operator.   
Nonetheless, the EGT admits the same block-diagonal form, and the regression problems are similar.

\section{Wavefront Measurement Error}

In Eq.~(\ref{Y=Hx}), both $\by_l$\ and $\bH_l$\ have terms that contain exponentiated residual phase, i.e., $\exp \big[ j \phi_\rr(\br,t) ]$.
The fact is that $\phi_\rr(\br,t) $\ must be somehow estimated from WFS datastream, and let $\hat{\phi_\rr}(\br,t)$\ represent this estimate.
Replacing $\phi_\rr(\br,t) $\ with $\hat{\phi_\rr}(\br,t)$\ in the regression both biases the coefficients and causes highly correlated uncertainties in their values.
This leads to a series of complex statistical issues that are introduced in [\citenum{Frazin16b}] and will the subject of future efforts by the author.

\section{Conclusions}

Standard differential imaging methods such as ADI and SDI have a variety of systematic errors and are particularly ineffective close to the star, where biggest returns on exoplanet science are expected.\cite{Stark_ExoEarthYield14,Brown_PlanetSearch15}.
The region within about $3 \lambda/D$\ (where $D$\ is the telescope diameter) is especially challenging due to the large angular displacement required for ADI, creating much incentive to explore methods beyond differential imaging.
Fortunately, millisecond focal plane telemetry is now becoming practical due to a new generation of near-IR detector arrays with sub-electron noise that are capable of kHz readout rates.
Combining these data with those simultaneously available from the WFS allows the possibility of self-consistently determining the optical aberrations (the cause of quasi-static speckles) and the planetary image.\cite{Frazin13,Frazin16a,Frazin16b}.   

Explicitly solving for the various aberrations in the optical system requires accounting for aberrations in various optical planes, leading to rather difficult and expensive computations, as many Fresnel propagations may be required.\cite{Frazin16a,Frazin16b}.
Here, an attractive alternative is presented.   
Instead of explicitly solving for the aberrations in a number of planes, one may instead assume that the diffraction problem can be solved by a Green's function with an unknown component that accounts for the aberration, including the non-common path component.
This unknown part of the Green's function is ``empirical Green's function,'' or EGF.
Here, it was shown that estimating the EGF leads to a straightforward set of regression equations that avoid the plane-by-plane propagations shown in [\citenum{Frazin16a,Frazin16b}].   
Furthermore, the EGF regressions can be made highly parallel due to their block-diagonal structure, which should make them practical.
It was also shown that the EGF can be generalized to treat unknown polarization effects by employing a similar ``empirical Green's Tensor'' (EGT).
   
The disadvantage of the EGF approach compared to explicit estimation the aberrations in multiple planes is that the EGF has many more degrees of freedom, although the richness of the millisecond datastreams may well afford the ability to estimate the larger number of parameters.   
It may be that as the field develops, the EGF (or EGT) would be most useful as an exploratory technique used guide and/or supplement techniques that solve for the aberrations explicitly.

\section*{Acknowledgments} 

This work has been supported by NSF Award \#1600138 to the University of Michigan.


\begin{thebibliography}{10}

\bibitem{Frazin13}
{Frazin}, R.~A., ``{Utilization of the Wavefront Sensor and Short-exposure
  Images for Simultaneous Estimation of Quasi-static Aberration and Exoplanet
  Intensity},'' {\em \apj}~{\bf 767},  21 (Apr. 2013).

\bibitem{Frazin14}
{Frazin}, R.~A., ``{Simultaneous ultra-high contrast imaging and determination
  of time-dependent, non-common path aberrations in the presence of detector
  noise},'' in [{\em Society of Photo-Optical Instrumentation Engineers (SPIE)
  Conference Series}{\nolinebreak\hspace{0.1em}]},  {\em Society of
  Photo-Optical Instrumentation Engineers (SPIE) Conference Series} {\bf 9145}
  (July 2014).

\bibitem{Frazin16a}
{Frazin}, R.~A., ``{Statistical framework for the utilization of simultaneous
  pupil plane and focal plane telemetry for exoplanet imaging. I. Accounting
  for aberrations in multiple planes},'' {\em Journal of the Optical Society of
  America A}~{\bf 33},  712 (Apr. 2016).

\bibitem{Frazin16b}
{Frazin}, R.~A., ``{Statistical framework for the utilization of simultaneous
  pupil plane and focal plane telemetry for exoplanet imaging. II. Implications
  of Wavefront Measurement Error for Regression Variables},'' {\em Journal of
  the Optical Society of America A, {\it in press}}  (2016).

\bibitem{Marois_SOSIE}
{Marois}, C., {Macintosh}, B., and {V{\'e}ran}, J.-P., ``{Exoplanet imaging
  with LOCI processing: photometry and astrometry with the new SOSIE
  pipeline},'' in [{\em Society of Photo-Optical Instrumentation Engineers
  (SPIE) Conference Series}{\nolinebreak\hspace{0.1em}]},  {\em Society of
  Photo-Optical Instrumentation Engineers (SPIE) Conference Series} {\bf 7736}
  (July 2010).

\bibitem{Rameau_ADI_SDI_limits15}
{Rameau}, J., {Chauvin}, G., {Lagrange}, A.-M., {Maire}, A.-L., {Boccaletti},
  A., and {Bonnefoy}, M., ``{Detection limits with spectral differential
  imaging data},'' {\em \aap}~{\bf 581},  A80 (Sept. 2015).

\bibitem{Mawet_SmallNumStatsSpeckle14}
{Mawet}, D., {Milli}, J., {Wahhaj}, Z., {Pelat}, D., {Absil}, O., {Delacroix},
  C., {Boccaletti}, A., {Kasper}, M., {Kenworthy}, M., {Marois}, C.,
  {Mennesson}, B., and {Pueyo}, L., ``{Fundamental Limitations of High Contrast
  Imaging Set by Small Sample Statistics},'' {\em \apj}~{\bf 792},  97 (Sept.
  2014).

\bibitem{Stark_ExoEarthYield14}
{Stark}, C.~C., {Roberge}, A., {Mandell}, A., {Clampin}, M., {Domagal-Goldman},
  S.~D., {McElwain}, M.~W., and {Stapelfeldt}, K.~R., ``{Lower Limits on
  Aperture Size for an ExoEarth Detecting Coronagraphic Mission},'' {\em
  \apj}~{\bf 808},  149 (Aug. 2015).

\bibitem{Brown_PlanetSearch15}
{Brown}, R.~A., ``{Science Parametrics for Missions to Search for Earth-like
  Exoplanets by Direct Imaging},'' {\em \apj}~{\bf 799},  87 (Jan. 2015).

\bibitem{SWIR_detector14}
{Fathipour}, V., {Memis}, O.~G., {Jang}, S.~J., {Khalid}, F., {Brown}, R.~L.,
  {Hassaninia}, I., {Gelfand}, R., and {Mohseni}, H., ``{Isolated nanoinjection
  photo detectors for high-speed and high-sensitivity single-photon
  detection},'' in [{\em Society of Photo-Optical Instrumentation Engineers
  (SPIE) Conference Series}{\nolinebreak\hspace{0.1em}]},  {\em Society of
  Photo-Optical Instrumentation Engineers (SPIE) Conference Series} {\bf 8868},
   3 (Sept. 2013).

\bibitem{SELEX_APD12}
{Finger}, G., {Baker}, I., {Alvarez}, D., {Ives}, D., {Mehrgan}, L., {Meyer},
  M., {Stegmeier}, J., {Thorne}, P., and {Weller}, H.~J., ``{Evaluation and
  optimization of NIR HgCdTe avalanche photodiode arrays for adaptive optics
  and interferometry},'' in [{\em Society of Photo-Optical Instrumentation
  Engineers (SPIE) Conference Series}{\nolinebreak\hspace{0.1em}]},  {\em
  Society of Photo-Optical Instrumentation Engineers (SPIE) Conference Series}
  {\bf 8453} (July 2012).

\bibitem{Saphira_eAPD14}
{Finger}, G., {Baker}, I., {Alvarez}, D., {Ives}, D., {Mehrgan}, L., {Meyer},
  M., {Stegmeier}, J., and {Weller}, H.~J., ``{SAPHIRA detector for infrared
  wavefront sensing},'' in [{\em Society of Photo-Optical Instrumentation
  Engineers (SPIE) Conference Series}{\nolinebreak\hspace{0.1em}]},  {\em
  Society of Photo-Optical Instrumentation Engineers (SPIE) Conference Series}
  {\bf 9148},  17 (Aug. 2014).

\bibitem{Mazin_MKIDS14}
{Mazin}, B., ``{MKIDs for TMT},'' in [{\em TMT in the Astronomical Landscape of
  the 2020s, Thirty Meter Telescope Science Forum, held 16-19 July, 2014 in
  Tucson Arizona. Online at:
  http://conference.ipac.caltech.edu/tmtsf2014/}{\nolinebreak\hspace{0.1em}]},
  (July 2014).

\bibitem{Hinkley_PDI09}
{Hinkley}, S., {Oppenheimer}, B.~R., {Soummer}, R., {Brenner}, D., {Graham},
  J.~R., {Perrin}, M.~D., {Sivaramakrishnan}, A., {Lloyd}, J.~P., {Roberts},
  Jr., L.~C., and {Kuhn}, J., ``{Speckle Suppression Through Dual Imaging
  Polarimetry, and a Ground-based Image of the HR 4796A Circumstellar Disk},''
  {\em \apj}~{\bf 701},  804--810 (Aug. 2009).

\bibitem{Hinnen_H2control}
Hinnen, K., Verhaegen, M., and Doelman, N., ``A data-driven $h_2$-optimal
  control approach for adaptive optics,'' {\em Control Systems Technology, IEEE
  Transactions on}~{\bf 16},  381--395 (May 2008).

\bibitem{Codona13}
{Codona}, J.~L. and {Kenworthy}, M., ``{Focal Plane Wavefront Sensing Using
  Residual Adaptive Optics Speckles},'' {\em \apj}~{\bf 767},  100 (Apr. 2013).

\bibitem{Clenet_aniso15}
{Cl{\'e}net}, Y., {Gendron}, E., {Gratadour}, D., {Rousset}, G., and {Vidal},
  F., ``{Anisoplanatism effect on the E-ELT SCAO point spread function. A
  preserved coherent core across the field},'' {\em \aap}~{\bf 583},  A102
  (Nov. 2015).

\bibitem{Breckinridge15}
{Breckinridge}, J.~B., {Lam}, W.~S.~T., and {Chipman}, R.~A., ``{Polarization
  Aberrations in Astronomical Telescopes: The Point Spread Function},'' {\em
  \pasp}~{\bf 127},  445--468 (May 2015).

\end{thebibliography}
\end{document}